\documentclass[sn-mathphys,Numbered, iicol]{sn-jnl}

\usepackage{graphicx}%
\usepackage{multirow}%
\usepackage{amsmath,amssymb,amsfonts}%
\usepackage{amsthm}%
\usepackage{mathrsfs}%
\usepackage[title]{appendix}%
\usepackage{xcolor}%
\usepackage{textcomp}%
\usepackage{manyfoot}%
\usepackage{booktabs}%
\usepackage{algorithm}%
\usepackage{algorithmicx}%
\usepackage{algpseudocode}%
\usepackage{listings}%

\usepackage[T1]{fontenc}

\usepackage{hyperref}
\usepackage{float}

\usepackage[normalem]{ulem}
\usepackage{pifont}

\usepackage{url}
\usepackage{breakurl}

\theoremstyle{thmstyleone}%
\theoremstyle{thmstyletwo}%

\theoremstyle{thmstylethree}%

\begin{document}

\title[Article Title]{Creating Valid Adversarial Examples of Malware}


\author*[1]{\fnm{Matou\v{s}} \sur{Koz\'{a}k}}\email{matous.kozak@fit.cvut.cz}

\author[1]{\fnm{Martin} \sur{Jure\v{c}ek}}\email{martin.jurecek@fit.cvut.cz}

\author[3]{\fnm{Mark} \sur{Stamp}}\email{mark.stamp@sjsu.edu}

\author[3]{\fnm{Fabio} \sur{Di Troia}}\email{fabio.ditroia@sjsu.edu}

\affil*[1]{\orgdiv{Faculty of Information Technology}, \orgname{Czech Technical University in Prague},\\ \orgaddress{\city{Prague}, \country{Czechia}}}

\affil[3]{\orgdiv{Department of Computer Science}, \orgname{San Jose State University}, \orgaddress{\city{San Jose}, \state{California}, \country{USA}}}



\abstract{Machine learning is becoming increasingly popular as a go-to approach for many tasks due to its world-class results. As a result, antivirus developers are incorporating machine learning models into their products. While these models improve malware detection capabilities, they also carry the disadvantage of being susceptible to adversarial attacks. Although this vulnerability has been demonstrated for many models in white-box settings, a black-box attack is more applicable in practice for the domain of malware detection. We present a generator of adversarial malware examples using reinforcement learning algorithms. The reinforcement learning agents utilize a set of functionality-preserving modifications, thus creating valid adversarial examples. Using the proximal policy optimization (PPO) algorithm, we achieved an evasion rate of 53.84\% against the gradient-boosted decision tree (GBDT) model. The PPO agent previously trained against the GBDT classifier scored an evasion rate of 11.41\% against the neural network-based classifier MalConv and an average evasion rate of 2.31\% against top antivirus programs. Furthermore, we discovered that random application of our functionality-preserving portable executable modifications successfully evades leading antivirus engines, with an average evasion rate of 11.65\%. These findings indicate that machine learning-based models used in malware detection systems are vulnerable to adversarial attacks and that better safeguards need to be taken to protect these systems.}




\maketitle


\section{Introduction} \label{sec:introduction}
\textbf{Mal}icious soft\textbf{ware}, also known as \textbf{malware}, conducts unwanted actions on infected systems. Protection of our devices is paramount as more and more of our lives are in the digital world. Cybersecurity professionals are developing new defense mechanisms to improve the detection capabilities of their antivirus (AV) programs. However, their opponents are advancing at the same, if not faster, rate, making the problem of malware detection a never-ending battle between attackers and antivirus developers.

According to the AV-TEST institute, more than 450,000 new malware samples are registered daily, totaling more than 150,000,000 new malicious programs in 2021 \cite{av_2021}. Nowadays, attackers are not focusing only on Windows devices, but other platforms such as Linux, Mac or Android are also targeted. However, Windows remains the go-to target for most attackers \cite{sophos2022}.

Using malware detection models based on machine learning (ML) yields promising results \cite{ucci2019survey}. Nonetheless, ML models are susceptible to adversarial examples (AEs) that can mislead the models \cite{papernot2016limitations}. For example, a minor modification of a malware file can make its feature vector resemble some of the feature vectors of benign files. Consequently, this can cause the malware classifier to make an incorrect prediction.

State-of-the-art antivirus programs incorporate both static and dynamic analysis in their inner workings. Static analysis methods, which do not require the executable to be run, provide us with information such as opcode sequences, program format information, and others. On the other hand, the dynamic analysis consists of studying the program's activity during execution and recording information such as API calls, registry and memory changes, and more \cite{damodaran2017comparison}.

In this work, we target our attack on static malware analysis for numerous reasons. Firstly, static detection is time-efficient as it does not involve the execution of binary executables. As a result, it is typically the first line of defense against unwanted threats and is thus a critical part of any antivirus program. Secondly, dynamic analysis requires executing malware inside a secure sandbox and recording its behavior, which is both time and technically demanding. Thirdly, malware authors can incorporate sandbox-evading techniques to detect that their malware is running inside a controlled environment and stop its malicious behavior \cite{erko2022malware, yuceel2022}.

Our goal is to implement a technique of adversarial attack at the level of samples, i.e., a technique that would create functional AEs. This task is considerably more demanding, as typical machine learning models operate at the level of feature vector, and reliable reverse mapping from a feature vector back to a binary file is difficult to perform. Therefore, we chose to perform adversarial perturbations on the binary level. To verify the functional preservation of our adversarial perturbations, we present a method comparing behavior patterns before and after modifications, which ensures maximum validity of the generated AEs.

Our adversarial attack works in a black-box scenario, mimicking the most difficult scenario where no information about the target classifier apart from the final prediction label is known to the attacker. While the defense methods should be tested in the most open (white-box) scenario to ensure the protection of the system in all settings \cite{kerckhoffs1883cryptographie}, we believe that the attack methods should be tested in the black-box scenario to provide a worst-case estimate of their abilities.

In this work, we introduce a set of functionality-preserving binary file modifications. Further, we train reinforcement learning agents to use these modifications to modify Windows malware binaries in order to avoid detection by the target classifier. Additionally, we compare trained and random agents and assess the transferability of our attacks to other malware detectors.

\paragraph*{Paper outline:}
\begin{itemize}
	\item In Section \ref{sec:background}, we establish the necessary background. Starting with an introduction to adversarial machine learning, continuing with a brief dive into reinforcement learning and finishing with a description of the portable executable file format.
	\item In Section \ref{sec:related_work}, we display related work focusing on the area of adversarial malware generation.
	\item In Section \ref{sec:proposed_method}, we define our method in detail. From modification of binary files and protocol to guarantee they retain their original functionality, to describing our reinforcement learning environment and agents.
	\item In Section \ref{sec:evaluation}, we introduce our experiments and present the results achieved.
	\item In \nameref{sec:conclusion}, we summarize the contributions of our work and suggest ideas for future research.
\end{itemize}


\section{Background} \label{sec:background}
In this section, we outline the necessary background to comprehend this paper. Firstly, we briefly introduce adversarial machine learning. Then we follow by describing the fundamental principles of reinforcement learning, and we finish by describing the portable executable file format in detail.

\subsection{Adversarial Machine Learning} \label{sec:adversarial_ml}
\textit{Adversarial machine learning} is an area of machine learning focusing on improving ML systems to withstand adversarial attacks both from inside (data poisoning) and outside (evasion attacks). An adversarial attack is a carefully created action to mislead the ML model. The victim model is also called a target model, and the attacker is called an adversary. Nevertheless, both attacker and adversary are used interchangeably in the current literature. The object responsible for misleading the target model is referred to as an \textit{adversarial example (AE)}. Adversarial machine learning is commonly employed in the malware detection domain to mislead AV products into incorrectly classifying malicious files as benign.

The success of an adversarial attack is dependent on the available knowledge of the targeted system \cite{huang2011adversarial}. When the adversary has access to the system and can examine its internal settings or training datasets, we call this a \textit{white-box} scenario. Contrarily, we refer to a situation as a \textit{black-box} scenario if the adversary only has access to a limited amount of information, typically just the model's final prediction, such as the malware/benign label for each sample that is presented. In between these two is a \textit{grey-box} scenario where the adversary has higher access to the system than in the black-box scenario, but the access is still limited to some parts. For example, the attacker can use the model's score or feature space but cannot access and modify its training dataset. Since the specific structure of the AV is typically unknown to the attacker, the black-box scenario is the most realistic for producing adversarial malware examples.

\subsection{Reinforcement Learning} \label{sec:reinforcement_learning}
\textbf{Reinforcement learning (RL)} is a branch of machine learning where an \textit{agent} equipped with a set of actions is learning how to reach its goal. The agent can be a bot learning to play a computer game or a physical robot working in a factory. Based on trial and error and appropriate feedback from an interactive element called the \textit{environment}, the agent learns which actions are ``good'' and ``bad'' for achieving its goal. The crucial challenge for reinforcement learning is a balance of \textit{exploration} and \textit{exploitation}, i.e., how to explore enough of the environment while maximizing the reward and reaching its goal. This section is based on the book \cite{sutton2018reinforcement}, where you can find more details and examples on this topic.

There are three key elements of reinforcement learning: the agent's policy, the reward signal, and the value function. A model of the environment is also included for some tasks but we will not specify this further.

The core part of any reinforcement learning agent is \textit{policy}. It is a function mapping from the states of the environment to an action from a set of agent actions representing the agent's behavior at a given time. If learned correctly, it should lead to a strategy that maximizes the total rewards the agent receives.

The \textit{reward signal} is an immediate response to a taken action provided by the environment. This signal grades action taken at a given state as good or bad concerning the agent's goal.

The \textit{value function} estimates how rewarding the current state is. The ultimate goal of every RL agent is to achieve the highest total reward, also called return. This goal usually cannot be accomplished by following states and actions with the highest immediate rewards but rather with the highest values, as these maximize the cumulative reward.

\begin{figure}[H]
	\centering	
	\includegraphics[width=0.8\linewidth]{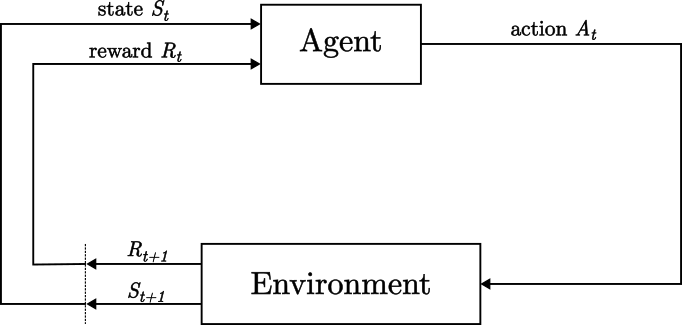}
	\caption{Interaction between the agent and the environment.}
	\label{fig:reinforcement_learning}
\end{figure}

Although there are other formal definitions of reinforcement learning, in this work, we follow the one presented in \cite{sutton2018reinforcement}. Reinforcement learning can be defined as repeated interactions between agent and environment at discrete time steps $t = 0, 1, 2, \ldots T$. At time step $t$, the environment is at a state $S_t \in S$ where $S$ is a set of all possible states. After the agent is presented with the state $S_t$, based on its policy $\pi$, creates a mapping $S_t \xrightarrow{} A_t \in A(S_t)$, where $A(S_t)$ is a set of all possible actions at state $S_t$. In many scenarios, $A(S_t)$ can change based on the current state $S_t$, but in others, it can remain fixed depending on the environment. After deciding on the action $A_t$, the chosen action is sent to the environment where it gets executed. The subsequent response from the environment gets presented to the agent in the form of a new state $S_{t+1}$, and reward $R_{t+1} \in R \subset{\mathbb{R}}$, where $R$ is the set of all possible rewards. Figure \ref{fig:reinforcement_learning} illustrates the interaction between the agent and the environment.

We call the exchange of actions, states and rewards between the agent and the environment across time steps $t = 0$ and $t = T$ an \textit{episode}. One episode can be characterized by the following sequence ending in the terminal state $S_T$, i.e., $S_0, A_0, R_1, S_1, A_1, \ldots, R_T, S_T$. Subsequently, the environment is reset, and a new independent episode begins.

As stated before, the agent's goal is to maximize the total of rewards  $G_t = R_{t+1} + R_{t+2} + \ldots + R_T$, also called expected \textit{return}. For the computation of expected return $G_t$, it is common to use a technique called discounting. This technique allows control over how far into the future agent should look, i.e., how much value it should assign to the future states. We calculate \textit{discounted return} as follows:

\begin{equation}
G_t = R_{t + 1} + \gamma R_{t + 2} + \gamma^2 R_{t + 3} + \cdots + \gamma^{T - t - 1} R_{T}
\end{equation}

\noindent
where $0 \leq \gamma \leq 1$ is a parameter called \textit{discount rate}. If $\gamma = 0$ agent only considers immediate reward, and the closer the $\gamma$ is to $1$, the higher value the agent gives to the future states. 

\subsubsection{Algorithms} \label{sec:rl_algorithms}
In this part, we briefly describe some of the algorithms popular in reinforcement learning, focusing on those we use later in this work. Detailed descriptions can be found in their original publications.

A popular algorithm called \textbf{Q-learning} was introduced in \cite{watkins1989learning} by Watkins. This algorithm works by iteratively estimating action-value function $Q(s, a)$, also called Q values using the following formula:

\begin{equation}
\scalebox{0.85}{$\begin{split}
Q^{new}(S_t, A_t) & = Q(S_t, A_t) \\ 
& + \alpha \Big( R_{t + 1} + \gamma \max_{a} \{ Q(S_{t + 1}, a) \} - Q(S_t, A_t) \Big)
\end{split}$}
\end{equation}

\noindent
where $a \in A(S_{t + 1})$. The $\max_{a} \{ Q(S_{t + 1}, a) \}$ represents the best value in the following state $S_{t+1}$. The calculated Q values are stored in a so-called Q table.

Improvement of the Q-learning algorithm called \textbf{deep Q-network (DQN)}, or deep Q-learning, was introduced by Mnih et al. \cite{mnih2013playing, mnih2015human}. DQN replaces the tabular manner of storing all state-action pairs $Q(s, a)$ with a function, usually taking the form of a neural network. The Q value is then defined as $Q(s, a; \xi)$, where $\xi$ can be one or more function parameters. 

The above-introduced algorithms learn the value functions and choose appropriate actions based on them. \textbf{Policy gradient (PG)} methods optimize the policy $\pi$ directly \cite{sutton1999policy}. The policy is parameterized by weight vector $\theta \in \mathbb{R}^n$. The decision on choosing action $a$ is therefore not only conditioned by the state $s$ but also by the vector $\theta$. The policy can then be defined as $\pi (a|s, \theta)$, i.e., the probability of taking action $a$ given that the agent is in the state $s$ with weight vector $\theta$ at time step $t$. Policy gradient methods use a gradient ascent algorithm to find the optimal value of $\theta$ to maximize the total return. 

\textbf{Proximal policy optimization (PPO)} is a variant of the policy gradient methods where the policy vector is updated only after several gradient ascent iterations \cite{schulman2017proximal}. This is in contrast with the original PG methods, which perform one gradient ascent update of the target policy per sample. As stated in the original paper, this improvement is easy-to-implement, yet it brings substantial performance improvements. 

\subsection{Portable Executable File Format} \label{sec:portable_executable}
\textbf{Portable executable (PE)} is a file format commonly found on Windows operating systems for various types of files, such as executables (EXEs) or dynamically linked libraries (DLLs). This file format is based on the Common Object File Format (COFF) found on Unix operating systems. It contains all the necessary information for the operating system (OS) loader to correctly map the PE file to system memory \cite{kowalczykPE}. 

\begin{figure}[h]
	\centering	
	\includegraphics[width=0.5\linewidth]{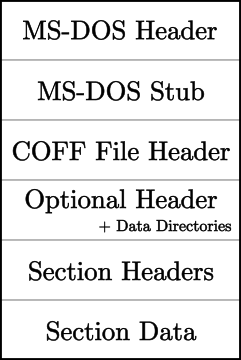}
	\caption{PE File Format}
	\label{fig:pe_file_format}
\end{figure}

In this section, we will describe the PE file format used for EXE files, as the usage of some fields differs from other file types. The PE file format has a rigid structure, as presented in Figure \ref{fig:pe_file_format}. Most of the information listed in this section comes from the official Windows documentation \cite{microsoftPE}.

\subsubsection{MS-DOS Header and Stub Program}
MS-DOS header and stub program are still part of the PE file format for backward compatibility with older operating systems (MS-DOS). Nowadays, if a modern Windows executable gets executed on MS-DOS, it should display some variation of the following message: ``This program cannot be run in DOS mode.''. 

\textbf{MS-DOS header} is 64 bytes long and is located at the beginning of the PE file. The first field of this header is \verb|e_magic|, also called a magic number. This field usually contains a value of \verb|0x5A4D|, a hexadecimal representation of the characters MZ, initials of one of the MS-DOS developers, Mark Zbikowski \cite{pietrekMicrosoftPE}. This field is followed by several other important fields for the MS-DOS system, which are not relevant for modern systems. The header is concluded with the \verb|e_lfanew| field, which stores a file offset to the COFF file header.

\textbf{MS-DOS stub program} is an actual valid MS-DOS program that would get executed on the MS-DOS operating system. This stub is located right after the MS-DOS header, and its size is variable depending on the program.

\subsubsection{COFF File Header}
Following the MS-DOS header and stub program is the \textbf{COFF file header}. It is located at the offset found in the \verb|e_lfanew| field from the MS-DOS header. Before the actual COFF header starts, there is a 4-byte field called \verb|Signature| that identifies the file as a PE file with a value of \verb|PE\0\0|. The following 20 bytes are the header itself, which contains general information regarding the PE file, such as \verb|NumberOfSection|, \verb|TimeDateStamp| or \verb|SizeOfOptionalHeader|.

\subsubsection{Optional Header}
Right after the COFF file header is located the \textbf{optional header}. Although it is called optional, for many files, such as EXEs, it is mandatory. It integrates core information for the OS loader. The header size can be found in the \verb|SizeOfOptionalHeader| field in the COFF header. This header has three main parts: standard fields, Windows-specific fields, and data directories.

The \textbf{standard fields} are eight fields used for each COFF file, which contain items such as:

\begin{itemize}
    \item \verb|Magic|: Indicates the type of optional header (32-bit/64-bit).
    \item \verb|SizeOfCode|: The size of the code section, usually called \textit{.text}.
    \item \verb|AddressOfEntryPoint|: Relative virtual address (RVA) of the entry point (the first program instruction when execution begins) after being loaded into memory.
\end{itemize}

The following are 21 fields belonging to the \textbf{Windows-specific fields} that contain unique information for the Windows operating system. Some of the fields are:

\begin{itemize}
    \item \verb|SectionAlignment|: Section alignment in memory.
    \item \verb|FileAlignment|: Section alignment on disk. The section is padded with zeros.
    \item \verb|SizeOfImage|: The size of the PE file. It must be rounded up to multiples of \verb|SectionAlignment|.
    \item \verb|SizeOfHeaders|: Sum of all header sizes rounded up to a multiple of \\ \verb|SectionAlignment|. 
    \item \verb|CheckSum|: Checksum value used to validate files such as drivers or DLLs.
    \item \verb|NumberOfRvaAndSizes|: Number of entries in data directories.
\end{itemize}

At the end of the optional header are placed the \textbf{data directories}. These directories form an array of 8-byte structures with two fields: RVA and size of the directory. There are 15 types of data directories, such as \textit{export}, \textit{import}, \textit{debug} or \textit{certificate} tables.
 
\subsubsection{Section Headers and Data}
Immediately following the optional header and data directories is the section table, also known as \textbf{section headers}. Each section header contains ten fields totaling 40 bytes in size. Examples of the fields include:

\begin{itemize}
	\item \verb|Name|: Eight bytes representing the name of the section padded with zeros, e.g., \verb|.text\0\0\0|.
	\item \verb|VirtualSize|: Section size when loaded into memory.
	\item \verb|VirtualAddress|: RVA of the first byte of the section. The headers must be sorted in ascending order by their corresponding virtual addresses. Additionally, the value of \verb|VirtualAddress| must be a multiple of \\ \verb|SectionAlignment|.
	\item \verb|SizeOfRawData|: The size of the section data on disk. It must be multiple of \verb|FileAlignment|. If the size is less than \verb|VirtualSize|, the rest of the section is padded with zeros.
	\item \verb|PointerToRawData|: Pointer to the beginning of the section on disk.
	\item \verb|Characteristics|: 4-byte flag indicating section attributes, e.g., the section contains executable code or can be shared in memory.
\end{itemize}

The location and size of the relevant \textbf{section data} are indicated in the corresponding section header. For EXEs, section data must be aligned using the \verb|FileAlignment| value found in the Windows-specific fields in the optional header. An ordinary PE file usually has several commonly found sections \cite{kowalczykPE}. Although their names may vary from file to file, their intentions remain the same. The most notable is the \textit{.text} section, which encapsulates all pieces of code. Naturally, the \verb|AddressOfEntryPoint| points to this section and also marks the end of the import address table, which is also part of this section. Additionally, the \textit{.bss}, \textit{.rdata}, and \textit{.data} sections store various types of data needed for the program to run.

One of the critical parts of nearly every EXE is the import directory table (IDT), or just the import table. This table is usually stored in the \textit{.idata} section, with each entry representing one of the imported DLLs. Since the size of the table is not fixed, the last entry is set to zero to indicate the end. Each entry contains the RVAs of the name of the imported DLL and the import address and lookup tables associated with this DLL. The import lookup table (ILT) is a table of imported function names from a given DLL. While stored on disk, the import address table (IAT) has the same structure and content as ILT. However, after the PE file is loaded into memory, the IAT entries contain the addresses of the imported functions instead of the function names.

The section name \textit{.debug} refers to a section containing debugging information. Five types of debugging information can be stored, each with a unique header structure. This section is not memory-mapped by default, and the PE file format also allows storing the information in a separate debug file.

When a PE file has a certificate, e.g., to ensure file origin or immutability, the location is specified in a security data directory inside the optional header. The security data directory points to the beginning of the attribute certificate table, which contains 20-byte entries for each certificate. This certificate table is usually not stored in one of the mapped sections but is appended to the end of the file in a segment typically called an \textit{overlay}.


\section{Related Work} \label{sec:related_work}
This section summarizes related publications that focus on creating adversarial malware examples. We break down this section into several subsections depending on the approach used to generate AEs. We begin by describing works based on the same technique as our work, i.e., reinforcement learning attacks. Then we present the researches that exploit the back-propagation algorithm commonly used in training deep neural networks with so-called gradient-based attacks \cite{goodfellow2015explaining, papernot2016limitations}. Note that most of these gradient-based attacks classify as white-box attacks since they work directly with the inner configurations of targeted models. Lastly, we mention several publications related to adversarial malware attacks that do not fit within these two categories. A summary of publications related to generating evasive AEs is presented in Table \ref{table:related_work}.

\begin{table*}[ht]
    \caption{Summary of related adversarial attacks. The symbol \ding{53} denotes that functionality preservation was not empirically verified.}
    \resizebox{\linewidth}{!}{\begin{minipage}{1.13\linewidth}
        \centering

        \begin{tabular}{@{}l|c|c|c|c@{}}
        \toprule
            Paper  & Knowledge & Problem Space & Target Model & Functionality \\ \midrule
            Anderson et al. \cite{anderson2018learning}        & grey-scale, black-box & binary files       & GBDT                          & \ding{53}                         \\
            Fang et al. \cite{fang2020deepdetectnet}           & black-box             & binary files       & deep neural network           & \ding{53}                         \\
            Song et al. \cite{song2020mab}                     & black-box             & binary files       & GBDT, MalConv, AVs            & sandbox                  \\
            Quertier et al. \cite{quertier2022merlin}          & grey-scale, black-box & binary files       & GBDT, MalConv, AVs, Grayscale & sandbox                  \\
            Kolosnjaji et al. \cite{kolosnjaji2018adversarial} & white-box             & binary files       & MalConv                       & \ding{53}                         \\
            Kreuk et al. \cite{kreuk2019deceiving}             & white-box             & binary files       & MalConv                       & \ding{53}                         \\
            Demetrio et al. \cite{demetrio2019explaining}      & white-box             & binary files       & MalConv                       & \ding{53}                         \\
            Yang et al. \cite{yang2021deepmal}                 & white-box             & binary files       & convolution neural network    & manually                 \\
            Hu and Tan \cite{hu2017generating}                 & white-box, black-box  & extracted features & deep neural network           & \ding{53}                         \\
            Ebrahimi et al. \cite{ebrahimi2020binary}          & black-box             & binary files       & MalConv                       & sandbox                  \\
            Demetrio et al. \cite{demetrio2021functionality}   & grey-box, black-box   & binary files       & GBDT, MalConv                 &  \ding{53}                        \\ \bottomrule
        \end{tabular}
    \end{minipage}}
    \label{table:related_work}

\end{table*}

\subsection{Reinforcement Learning-Based Attacks}
One of the first works done in the domain of generating AEs using reinforcement learning was published in 2018 by Anderson et al. \cite{anderson2018learning}. The authors presented a gym-malware framework equipped with RL agents and an OpenAI gym \cite{brockman2016openai} environment. They targeted a gradient-boosted decision tree (GBDT) trained on 100,000 binary files and achieved an evasion rate of up to 24\%, depending on the dataset used. 

In \cite{fang2020deepdetectnet}, Fang et al. presented two models, a detector called DeepDetectNet and a generator of AEs called RLAttackNet. In a pure black-box scenario, their generator, based on the DQN algorithm, bypassed their own detector in 19.13\% of cases. The authors later used the AEs to retrain their DeepDetectNet model, and the evasion rate of RLAttackNet dropped to 3.1\%. 

The MAB-Malware framework was presented by Song et al. \cite{song2020mab}. As an agent, the authors used a multi-armed bandit (MAB) model while adjusting the agent's action set in real-time by adding successful action-content pairs. Further, the authors introduced an action minimization procedure, which removes unnecessary modifications after successful evasion, reducing the final size of AEs. The authors targeted the GBDT by EMBER \cite{anderson2018EMBER}, MalConv \cite{raff2017malware}, and several commercial AV detectors. They recorded a high evasion rate of 74.4\% against GBDT, 97.7\% against MalConv, and up to 48.3\% against commercial AVs. Further, they validated the functionality preservation of generated AEs by comparing file signatures with genuine malware files.

Quertier et al. in \cite{quertier2022merlin} used reinforcement learning algorithms to attack MalConv, GBDT by EMBER and Grayscale (convolutional neural network interpreting PE binaries as images) classifiers in grey-scale settings with available prediction scores for learning. Further, the authors targeted commercial AV in a pure black-box environment as well. They used the DQN and REINFORCE (policy gradient algorithm) agents and achieved a high evasion rate against all targeted models, including an 80\% evasion rate with REINFORCE against GBDT, a 100\% perfect evasion against MalConv with both algorithms and a 70\% evasion rate against commercial AV with REINFORCE. However, Quertier et al. did not specify what commercial AV they were targeting, nor did the authors disclose their models for further use.

\subsection{Gradient-Based Attacks}
In \cite{kolosnjaji2018adversarial}, Kolosnjaji et al. proposed a gradient-based attack against the MalConv malware detector \cite{raff2017malware}. Their attack only targeted the overlay part of the file and achieved an evasion rate of 60\% while modifying less than 1\% of total bytes.

Kreuk et al. \cite{kreuk2019deceiving} used a gradient-based attack, limited to injecting small-scale chunks of bytes into unused parts or at the end of the file. The authors argued that these modifications do not change the functionality of the file but without further justification. The authors scored a high evasion rate of 99\% against the MalConv classifier.

Another attack on MalConv was carried out by Demetrio et al. in \cite{demetrio2019explaining}. Using an integrated gradient method, the authors studied which sections of binaries stimulate the MalConv classifier and thus are vulnerable to adversarial attacks. They found that MalConv partially bases its prediction on features in the DOS header and achieved an evasion rate of more than 86\% by only modifying the DOS header.

A variation on the method introduced in \cite{kreuk2019deceiving} was presented in 2021 by Yang et al. \cite{yang2021deepmal}. The authors treated the input EXEs as images, used as input into a convolution neural network. They calculated necessary byte perturbations for evasion and then transformed them into specific byte sequences. Depending on the location of the given perturbation, the resulting byte sequence was either a dead-code or API call instruction. The authors conducted a theoretical examination of the above-mentioned modifications to confirm that their modifications preserve functionality. Their introduced perturbations reduced the accuracy of several ML detectors by up to 94\%.

\subsection{Other Methods}
A generative adversarial network (GAN) called MalGan was proposed in \cite{hu2017generating} as a method of generating AEs. During training, the authors used a substitute detector in the form of a deep neural network and their results demonstrate high attack transferability between the substitute and target models, with near-perfect evasion against the random forest, decision tree, and linear regression algorithms. However, they worked in the feature space of extracted API calls and did not provide a method to convert the adversarial feature vectors back to real-world EXEs.

One of the few works that addresses the data poisoning issue is \cite{chen2018automated} by Chen et al. Even though this work is centered on the Android operating system, the results could also apply to Windows systems. The authors saw up to 30\% drops in accuracy after injecting their data while targeting pure ML models. As a defense mechanism, they introduced a camouflage detector that detects suspicious samples inside the training dataset and boosts the detector's accuracy by at least 15\%.

Ebrahimi et al. proposed a generative sequence-to-sequence language model in the form of a recurrent neural network \cite{ebrahimi2020binary}. This network was trained on benign binaries to generate adversarial benign bytes. These benign bytes were subsequently appended to malware EXEs to produce adversarial malware instances. The authors achieved an average evasion rate of 73.24\% against the MalConv classifier. Their findings suggest that increasing padding size increases the evasion rate but with diminishing returns as the value grows. The authors conducted a behavior analysis to validate that the functionality of malicious EXEs did not change after appending generated benign bytes.

In \cite{demetrio2021functionality}, Demetrio et al. presented a black-box attack named GAMMA. The GAMMA attack tackles the problem of creating AEs as an optimization problem, with the main requirement being maximum evasion and minimal inserted content size. The optimization problem was solved using a genetic algorithm that used the traditional selection process, cross-over, and mutation. In the training phase, the authors targeted the GBDT and MalConv classifiers to later attack real-world AVs hosted on the VirusTotal website, successfully bypassing 12 out of 70 detectors on average.


\section{Adversarial Malware Generator} \label{sec:proposed_method}
We introduce a complete framework for generating adversarial malware examples called AMG (Adversarial Malware Generator). AMG consists of a tested PE file modifier, which can be easily expended with additional modifications, an environment in the Open AI Gym format \cite{brockman2016openai} working with raw binary files and a set of optimized reinforcement learning agents ready to use.

Our approach falls into the category of evasive black-box attacks. In other words, we are performing an adversarial attack to mislead the target model (e.g., antivirus) to classify malware samples as benign. Our objective is to execute small modifications on PE files that do not alter the original functionality but can make them undetectable to the antivirus. Our attack targets static malware analysis, where the detector makes decisions without examining the EXE's behavior. We set our adversarial attack in a black-box scenario where only the target classifier's hard predictions (malware/benign) are known to the attacker, as this is the most difficult scenario for the attacker. In addition, we believe that using the black-box approach limits the likelihood of our attack being successful only against a specific classifier and consequently increases the potential of transferability to other detectors. However, additional research in this field should be conducted in the future.

In our work, we modified the existing framework called gym-malware by Anderson et al. \cite{anderson2018learning}, which provides an environment for training reinforcement learning agents on binary samples. We rewrote most parts because the existing code did not meet our vision and goals. In particular, they used the \verb|LIEF| \cite{LIEF} library for modifying PE files, whereas we used the \verb|pefile| \cite{carrera2017pefile} Python library. We found that the \verb|LIEF| library can make unnecessary changes to the original binary and that their modifications did not retain the same functionality as is shown later in Section \ref{subsec:evaluation_functionality_preservation}. In addition, in their training setup, the agent is presented with an observation space that coincides with the feature space of the target classifier as such it cannot be classified as a pure black-box setup. Furthermore, we think that using the feature space of the target classifier can detriment the transferability of trained agents to other detectors. For this reason, we used a different observation space that is not used by any of the classifiers we targeted. Nonetheless, the work by Anderson et al. \cite{anderson2018learning} is a key stepping stone for future research as it is one of the first complete frameworks for deploying RL agents for adversarial malware example generation.

In the following subsections, we describe in detail our proposed method, starting with the PE file modifications we use and how we validate them. Later, we introduce our RL environment setup and agents.

\subsection{PE File Modifications} \label{subsec:pe_file_modifications}
For implementing PE file modifications, we used the \verb|pefile| \cite{carrera2017pefile} Python library. This library provides a simple interface for accessing all parts of the PE file format, such as file and optional header fields or individual sections. The description of the PE file format can be found in Section \ref{sec:portable_executable}. We implemented various modifications of the binary files, all obeying the structure of the PE file format. While we had taken inspiration from state-of-the-art related works, such as the gym-malware mentioned above, we also introduced new modifications. In total, we implemented ten modifications which are described below:

\begin{itemize}
	\item \textit{Break CheckSum}: Set the \verb|CheckSum| field from the optional header to zero.

	\item \textit{Append to overlay}: Append a random benign content to the end of the file.

	\item \textit{Remove debug}: Clear the debug entry in the list of data directories and remove the respective debug information from the file.

	\item \textit{Remove certificate}: Clear the security entry in the list of data directories and remove the certificate data from the file.

	\item \textit{Add new section}: Add a new section to the PE file if possible. Firstly, it is necessary to check if the file has enough free space between the last section header and the beginning of section data (at least 40 bytes). If so, we can increase the file size and add a new section header and data. To preserve the original PE file structure as much as possible, we also move the old overlay data and, if present, redirect the security data directory to the new address.

	\item \textit{Append to section}: Append benign content to one of the existing sections if possible. First, we need to find a section with the possibility of adding extra content, i.e., the virtual size of the section is greater than its raw size. If we encounter one, we fill the empty space with benign content.

	\item \textit{Rename section}: Choose one section at random and rename it to one of the section names commonly used in benign files.

	\item \textit{Increase TimeDateStamp}: Increase the value of \verb|TimeDateStamp| in the COFF file header by 500 days\footnote{We picked 500 days because it is a considerable period of time and it is not a multiple of one year.\label{fn:500_days}}.

	\item \textit{Decrease TimeDateStamp}: Decrease the value of \verb|TimeDateStamp| in the COFF file header by 500 days\footref{fn:500_days}.

	\item \textit{Append new import}: Add a new section to the PE file with import data if possible. This process is similar to the preceding add new section modification with the only change that the section content is not random benign content but import data. If already present, we take the old IDT table from the PE file and append a new entry, a randomly chosen DLL. Then we prepare entries for imported functions stored in the IAT and ILT tables. All of this information is then stored in a newly added section, and the import data directory is pointed to the new IDT table.
\end{itemize}

\subsubsection{Validity of PE File Modifications} \label{subsec:functionality_preservation}
We believe that preserving the original functionality (i.e., validity) of executable binary files is a critical part of generating adversarial malware samples. Without emphasizing this criterion, we cannot guarantee that the resulting AE will still be a working executable with the same functionality as the original file. We have found that more than simply checking the syntax of the PE file format is needed to maintain functionality, so we designed the following protocol to test the validity of PE file modifications.

To ensure that the functionality of an executable after adversarial perturbations is as close as possible to the original file behavior, we used a Cuckoo Sandbox \cite{cuckooSandbox}. Cuckoo Sandbox is an open-source automated malware analysis tool that can run malicious files and examine their behavior. Even though it is predominantly intended for malware analysis, we also used it to analyze benign files as it provides behavioral analysis, which we utilized to track any changes in the functionality of executables. We decided to use benign files instead of malware EXEs for testing the modifications because malware authors can insert checks into their programs that monitor whether their malware is running in a sandbox environment and change its behavior accordingly \cite{erko2022malware, yuceel2022}. Consequently, by using benign files, we limit the possibility of artificial activity of the tested binaries, and thus we can better analyze the reported behavior.

In contrast with other approaches \cite{song2020mab,quertier2022merlin}, we do not verify the functionality preservation of generated AEs, but we propose validating each modification individually before the generation process. Therefore our approach is more time-efficient as it does not require discarding nonfunctional AEs during or at the end of the generation procedure. 

For our method, we selected a set of benign EXEs $\mathcal{D}$ that was executable in the sandbox environment and studied their respective behavior reports. Namely, we looked into three features found in the Cuckoo analysis report: signatures, API calls, and processes:

\begin{itemize}
	\item \textit{Signatures}: Predefined patterns that are used to compare with the examined file. They are used predominately for malware detection to cluster malware into their respective families. Nevertheless, they can also classify types of actions, such as file open/write or access to system files, which also occur with benign files.
	\item \textit{API calls}: Function calls by a program to external libraries during program execution.
	\item \textit{Processes}: Main process and sub-processes started by a program.
\end{itemize}

To combat the variability of results reported by the sandbox environment, we conducted three testing rounds and considered the feature matched if it got at least 95\% agreement between rounds. We selected the value of 95\% to allow a small margin of error and to get all unmodified files reliably matched.

The set of untampered benign files $\mathcal{D}$ is used as a control dataset to test whether the functionality of the PE file has changed after the modification $M$. Firstly, the modification $M$ is applied on each file $f \in \mathcal{D}$, creating a dataset of modified files $f^{M} \in \mathcal{D}^{M}$. Next, the modified dataset $\mathcal{D}^M$ is compared with the unmodified dataset $\mathcal{D}$ according to the same three features we mentioned earlier by performing three rounds of Cuckoo analysis. We consider the modification a failure if the modified file cannot be run in the Cuckoo Sandbox. If the file executes successfully, we compare the three generated test analysis reports with all three control reports. We look at each feature individually, matching it with each control file. The feature is considered matched if it has an agreement of at least 95\% with one of the control files. Overall, the modified file is considered successfully modified (i.e., the original functionality has been preserved) if it matches at least two of its features with control reports. More detailed pseudo-code on how we evaluate each modified file $f^M$ is shown in Algorithm \ref{alg:functionality_preservation}.

\begin{algorithm*}[h]
	\caption{Evaluation of file validity after modification.}
	\label{alg:functionality_preservation}
	\begin{algorithmic}[1]
    \State $f \gets$ original file
    \State $f^M \gets$ modified file
    \State $CRs, TRs \gets$ empty lists
    \item[]
    
	\For{$i \gets 1$ to 3} \Comment{Create lists of control ($CRs$) and test ($TRs$) reports} 
		\State $CRs_{i} \gets$ Cuckoo-analysis($f$)
		\State $TRs_{i} \gets$ Cuckoo-analysis($f^M$)
	\EndFor    
    \item[]

	\State $c \gets 0$ \Comment{Number of matched features}
	\For{$TR \in TRs$}
		\If{$TR$ is failure}
			\Return FAILURE
		\EndIf
		
		\For{$feature \in$ [signatures, API calls, processes]} 
			\For{$CR \in CRs$}
				\If{match-feature($feature$, $TR$, $CR$)}
					\State $c \gets c + 1$ \Comment{More than 95\% agreement between $TR$ and $CR$}
					\State \textbf{break}				
				\EndIf
			\EndFor	
		\EndFor
	\EndFor
    \item[]

    \If{$c >= 2$} 
    	\Return SUCCESS \Comment{At least two features are matched}
    \Else{}
    	\Return FAILURE
    \EndIf
	\end{algorithmic}
\end{algorithm*}

By testing only a single modification $M$ rather than resulting AEs spanning multiple changes, we can better focus on each modification and trace potential errors. In addition, our protocol is time-efficient because the evaluation can be performed before generating evasive AEs.

\subsection{Malware Environment}
As mentioned in Section \ref{sec:reinforcement_learning}, RL algorithms are based on learning through feedback provided by the environment. We worked with a commonly used environment format developed by the OpenAI company called Gym \cite{brockman2016openai}. The \verb|Gym| is an open-source Python library equipped with a standardized API for agent-environment interaction. The source codes of our implementation are available in the GitHub repository\footnote{\url{https://github.com/matouskozak/AMG}}.

A critical part of the environment is the target classifier, as each action is rewarded with respect to its predictions. We studied two ML classifiers, MalConv and GBDT, both publicly available on GitHub with pre-trained configurations\footnote{\url{https://github.com/endgameinc/malware_evasion_competition}}. MalConv is a deep convolutional neural network that does not require complex feature extraction procedures because it uses the entire EXE (truncated to 2,000,000 bytes) as an input feature vector \cite{raff2017malware}. On the other hand, GBDT is a gradient-boosted decision tree trained using the LightGBM framework that requires converting the input executable to an array of 2,381 float numbers \cite{anderson2018EMBER}. Note that we did not directly target the MalConv classifier but only used it to explore whether adversarial attacks can be transferred between ML classifiers.

\subsection{Reinforcement Learning Agents} \label{subsec:proposed_method_rl_agents}
In total, we experimented with three RL agents, deep Q-network (DQN), vanilla policy gradient (PG) and proximal policy optimization (PPO). We chose these reinforcement learning algorithms because they are well-known in the reinforcement learning community and represent both on-policy and off-policy approaches. A more detailed description of these algorithms can be found in the previous Section \ref{sec:rl_algorithms}. We made use of the implementations that were offered by the \verb|Ray RLLib| \cite{liang2017ray} reinforcement learning library.


\section{Evaluation} \label{sec:evaluation}
In this section, we first describe our experimental setup. Secondly, we evaluate the validity of various PE file modifiers. Next, we present how we optimized individual RL algorithms to achieve the highest possible evasion rate against the GBDT target classifier. Lastly, we evaluate how generated AEs transfer to other malware detectors. In real-world circumstances, the transferability of AEs between detectors is paramount, as the victim's defenses can be unknown to the attacker.

\subsection{Setup} \label{subsec:setup}
\textbf{Datasets}: We use two datasets. A dataset of benign binaries, including more than 4,000 executables, was scrapped from the fresh Windows 10 installation. These benign files are only used for testing the preservation of functionality after modification, as mentioned later in Section \ref{subsec:functionality_preservation}. Second, a dataset of malware files was obtained from the VirusShare\footnote{\url{https://virusshare.com/}} repository. In total, we operate with 7,000 malware files divided into three parts: a training dataset consisting of 4,000 files, a validation set of 1,000 samples and a testing set containing 2,000 files.

\noindent \textbf{Evasion rate}: The principal metric we use in this work is called an evasion rate. This metric denotes the ratio of misclassified files by the target classifier and is calculated as follows:

\begin{equation}
	evasion\ rate = \dfrac{\#\ missclassified}{total} \cdot 100\%
\end{equation}

\noindent
where $total$ stands for the total number of files submitted to the target classifier after discarding files that were already incorrectly predicted before modification.

\noindent \textbf{Computer setup}: Our experiments were executed on a single computer platform with two server CPUs (Intel Xeon Gold 6136, base frequency 3.0Ghz, 12 cores), one GPU (Nvidia Tesla P100, 12 GB of video RAM) and 754 GB of RAM running the Ubuntu 20.04.5 LTS operating system.

\subsection{Evaluation of the Preservation of Functionality} \label{subsec:evaluation_functionality_preservation}
We used a set of 100 benign EXEs, complying with the requirements introduced in Section \ref{subsec:functionality_preservation}, to evaluate our PE file modifications described in Section \ref{subsec:pe_file_modifications}, and to compare them with several PE file modifiers from well-known frameworks for generating adversarial malware examples. Namely, we tested gym-malware\footnote{\url{https://github.com/endgameinc/gym-malware}}, Pesidious\footnote{\url{https://github.com/CyberForce/Pesidious}} and MAB-Malware\footnote{\url{https://github.com/weisong-ucr/MAB-malware}} generators. Both gym-malware and Pesidous use the \verb|LIEF| library for modifying binaries, whereas MAB-Malware, the same as our approach, uses the \verb|pefile| library. Additionally, Pesidous uses the \verb|PE Bliss| \cite{peBliss} C++ library for rebuilding PE files. We chose these frameworks because they are all publicly available on GitHub and share the reinforcement learning approach with our work.

\begin{table}[h]
	\centering
	\caption{Number of valid files after modification out of a total of 100.}
    \resizebox{\columnwidth}{!}{\begin{minipage}{\textwidth}
		\begin{tabular}{@{}l|cccc@{}}
\toprule
action         			& gym-malware 	& Pesidious		& MAB-malware 	& \textbf{AMG} 	\\ \midrule
break checksum     		& 89     		& \ding{53}    	& 100			& 100   	\\
create new entry point 	& 17     		& \ding{53}    	& \ding{53} 	& \ding{53} \\
append new import   	& 20     		& 42    	  	& \ding{53} 	& 66   		\\
overlay append     		& 100     		& 99    	  	& 100			& 100  		\\
remove debug      		& 90     		& \ding{53}    	& 100			& 100   	\\
remove certificate   	& 22     		& \ding{53}    	& 90			& 91   		\\
add new section   		& 4      		& 85    	  	& 75			& 98   		\\
append to section   	& 8      		& \ding{53}    	& 99   			& 99 		\\
rename section     		& 89     		& 89     	  	& 99			& 100   	\\
upx pack        		& 73     		& \ding{53}    	& \ding{53} 	& \ding{53} \\
upx unpack       		& 100     		& \ding{53}    	& \ding{53}		& \ding{53} \\
increase TimeDateStamp 	& \ding{53}  	& \ding{53} 	& \ding{53}		& 100		\\
decrease TimeDateStamp 	& \ding{53}  	& \ding{53}	  	& \ding{53}		& 100		\\ 
\bottomrule
		\end{tabular}
	\end{minipage}}
	\label{table:pe_modifications_cuckoo}
\end{table}

We evaluated each PE file modification in an isolated setting, i.e., in each run of Algorithm \ref{alg:functionality_preservation}, the input file was perturbed by a single modification. We present the results of functionality preservation testing in Table \ref{table:pe_modifications_cuckoo}, where the first column represents different PE modifications, and the following ones represent the number of valid files for PE modifiers from the respective frameworks. The symbol \ding{53} denotes that the operation was not implemented by the framework. Apart from the modifications mentioned in Table \ref{table:pe_modifications_cuckoo}, authors of MAB-Malware also implemented code randomization operation. However, we could not reproduce the code locally for our dataset, so we did not include it in our testing. 

We can see that the AMG modifications equaled or surpassed all other tested frameworks. In contrast, the gym-malware framework recorded the worst results with some of the modifications, such as adding a new section or appending to a section created valid binaries in only 4 and 8 cases, respectively. Overall, we can see that MAB-Malware and AMG, the PE file modifiers that use the \verb|pefile| library, better preserve the original functionality than the PE modifiers using the \verb|LIEF| library, such as gym-malware and Pesidious.

\subsection{Generating Adversarial Malware Examples}\label{subsec:generating_AMG}
In the following experiment, we focus on optimizing our generator of malicious AEs (AMG) against the GBDT classifier. We define the following procedure used for each RL algorithm. The first step is finding the optimal maximal number of modifications for the RL algorithm. For this part of the experiment, we leave the agent's parameters at their default settings as set in the \verb|Ray RLLib|. The range we test is between 5 and 200 modifications, and we choose the optimal value based on two criteria. Firstly, we try to maximize the evasion rate achieved by the agent and secondly, we try to minimize the increased size of evasive AEs.

After determining the maximum number of modifications, we conduct a hyperparameter search using the grid search method over two hyperparameters, the learning rate (lr, $\alpha$) and the discount rate (gamma, $\gamma$), leaving the rest of the parameters at the default settings as defined by the authors of \verb|Ray RLLib|. Based on the highest mean episode reward (mean of all rewards received during a single episode) recorded during 100 training iterations, we select the best four agent configurations and let them train for another 900 iterations. After the training finishes, we test these agents on the validation set and determine the best agent configuration for the RL algorithm. 

\begin{figure*}[h]
	\centering	
	\includegraphics[width=\linewidth]{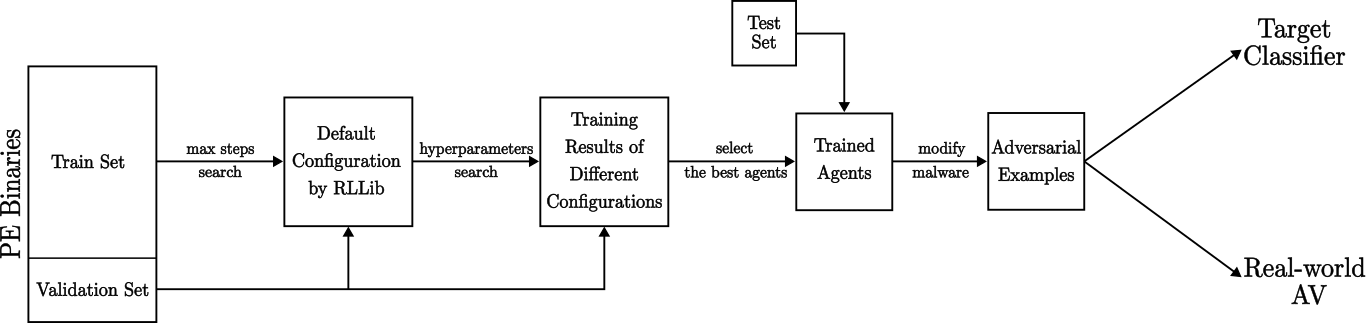}
	\caption{Workflow of our training and testing procedure for generating adversarial malware examples.}
	\label{fig:workflow_protocol}
\end{figure*}

Subsequently, we introduce our testing dataset, which is presented to the final RL agent. The obtained results are then used to compare different RL algorithms and to verify the success of the training stage. The complete overview of our optimization and evaluation workflow can be found in Figure \ref{fig:workflow_protocol}.

\subsubsection{Optimization and Selection of RL Agents}
Firstly, we followed the optimization process described earlier to tune the maximum number of steps, learning rate, and discount rate hyperparameters. Based on the results achieved by the respective RL algorithms on the validation set, we selected the following configurations listed in Table \ref{table:RL_validation_results}. 

\begin{table}[h]
	\centering
	\caption{The best configuration of each tested RL algorithm and their results against the GBDT classifier on the validation set.}
	\resizebox{\columnwidth}{!}{\begin{minipage}{\textwidth}
\begin{tabular}{@{}llll|ll@{}}
\toprule
agent & max steps & $\alpha$ & $\gamma$ & evasion rate [\%] & size increase [\%]\\ \midrule
DQN & 50        & 0.01    & 0.5      & 79.93        & 4.64          \\
PG  & 20        & 0.01    & 0.75     & 69.33        & \textbf{3.24}          \\
PPO & 50        & 0.0001  & 0.5      & \textbf{90.38}        & 9.2           \\ \bottomrule
		\end{tabular}
	\end{minipage}}
	\label{table:RL_validation_results}
\end{table}

The values reported show that PPO outperformed the other RL algorithm tested, although it was offset by a 9.2\% size increase. Overall, the PG method performed the poorest, as it did not exceed 70\% evasion rate, but it increased the resulting AEs the least, by only 3.24\% on average.

\subsubsection{Evaluation of RL Agents}
Following that, we used our test dataset to evaluate the best configuration of each RL algorithm from Table \ref{table:RL_validation_results}. Furthermore, we included a so-called random agent in the following results. This agent represents a repeated random application of the modifications described in Section \ref{subsec:pe_file_modifications} until the GBDT classifier is bypassed or 50 alterations are reached. Including the random agent allows us to better understand the learning capabilities of the trained RL algorithms.

\begin{table}[h]
	\centering
	\caption{Results of the best configuration of each tested RL algorithm and random agent against the GBDT classifier on the test set.}
		\begin{tabular}{@{}l|cc@{}}
\toprule
agent  & evasion rate [\%] & size increase [\%] \\ \midrule
DQN    & 52.72         & 4.63                   \\
PG     & 46.75         & 3.11                   \\
PPO    & \textbf{53.84} & 3.57                  \\
random & 36.88         & \textbf{1.07}          \\ 
\bottomrule
		\end{tabular}
	\label{table:RL_test_results}
\end{table}

The test results are shown in Table \ref{table:RL_test_results}. We immediately see significant performance decreases for all agents compared to the validation set, suggesting possible overfitting. However, the agents still achieved decent results by overcoming GBDT detector in approximately half of the cases. The trend is similar to what we saw on the validation set, with PPO recording the highest evasion rate of 53.84\% and the PG agent recording the lowest. The measured results represent an improvement of up to 17\% over the random agent, which, on the other hand, had the lowest average size increase of 1.07\%. To summarize our results, the overall best RL algorithm against the GBDT classifier is PPO with $\gamma=0.5$ and $\alpha=0.0001$, striking the highest evasion rate while maintaining a reasonable size increase of generated AEs.

\subsection{Transferability of Adversarial Attack} \label{subsec:transferability_AA}
In the previous experiment, we considered the GBDT classifier as our target model. In the final evaluation phase, we test the transferability of adversarial attacks between GBDT and other malware detectors. Based on the functionality preservation results from Table \ref{table:pe_modifications_cuckoo}, we included the MAB-Malware generator in our transferability experiment as it obtained a similar validity of modified files as our AMG framework and targets the GBDT classifier as well.

\subsubsection{Transferability to the MalConv Classifier}
To begin, we examine the transferability of adversarial attacks from GBDT to the MalConv classifier. We employed the best-trained AMG agent (PPO with $\gamma=0.5$ and $\alpha=0.0001$), the random agent with the AMG modifications, and MAB-Malware to generate AEs against the GBDT detector and then test them against the MalConv classifier.

\begin{table}[h]
	\centering
	\caption{Transferability of adversarial attacks targeted against GBDT to MalConv.}
	\begin{tabular}{@{}l|cc@{}}
\toprule
            & GBDT & MalConv \\ \midrule
MAB-Malware   & \textbf{76.12} & \textbf{60.1}    \\
AMG-PPO     & 53.84      & 11.41   \\
AMG-random  & 36.88      & 7.65    \\ \bottomrule
	\end{tabular}
	\label{table:transferability_GBDT_MalConv}
\end{table}

The recorded evasion rates against the GBDT and MalConv classifiers are listed in Table \ref{table:transferability_GBDT_MalConv}. From these results, we can conclude that MAB-Malware outperforms our AMG agents in terms of GBDT evasion as well as transferability to the MalConv classifier. The MAB-Malware recorded evasion rates of 76.12\% and 60.1\% against the GBDT and MalConv models, respectively. In comparison, the AMG framework struggled to transfer the performance recorded against GBDT to MalConv, with a considerable decrease in performance from 53.84\% down to 11.41\%. 

\subsubsection{Transferability to Real-World Antiviruses}
In the final stage of our experiments, we compare the performance of the above-mentioned generators of adversarial malware against leading AV programs. The transferability of generated AEs from lightweight ML detectors to commercially deployed AVs is an essential feature that AE generators should have, as developing AEs against AVs is time-consuming and could reveal the attacker.

We conduct this assessment on a selection of antivirus programs based on the March 2023 antivirus comparative study by the Austrian AV testing laboratory AV-Comparatives \cite{av-comparatives_2023}. We used the VirusTotal\footnote{\url{https://www.virustotal.com/}} website as a substitute for local instances of selected AV engines.

\begin{table*}[ht!]
	\centering
	\caption{Transferability of adversarial attacks targeted against GBDT to AVs.}
	\begin{tabular}{@{}l|llllllll|l@{}}
\toprule
            & AV-1 & AV-2 & AV-3 & AV-4 & AV-5 & AV-6 & AV-7 & AV-8 & average \\ \midrule
MAB-Malware & 1.22       & 0.41  & 1.66  & 1.9         & 1.53  & 3.0       & 7.76   & 3.38     & 2.61    \\
AMG-PPO     & 2.39       & 0.41  & 2.84  & 2.75        & 1.79  & 2.41      & 3.96   & 1.9      & 2.31    \\
AMG-random  & \textbf{9.37}  & \textbf{2.15} & \textbf{9.86} & \textbf{11.74} & \textbf{9.22}  & \textbf{12.88} & \textbf{34.48} & \textbf{3.54} & \textbf{11.65} \\ \bottomrule
	\end{tabular}
	\label{table:transferability_GBDT_AVs}
\end{table*}

The recorded evasion rates of generated AEs against AV detectors are presented in Table \ref{table:transferability_GBDT_AVs}. We anonymized the recorded results to minimize the possible risk of misuse of our work and to comply with the VirusTotal policies. At first glance, we can see that AMG-PPO and MAB-Malware decreased their performance to 2.31\% and 2.61\%, respectively. This represents a significant drop in evasion rates compared to the original ones recorded against the targeted GBDT classifier. This decrease could indicate that employing the GBDT classifier as a substitute model for generating malicious AEs does not lead to successful evasion against real-world AV programs and that better surrogate models should be utilized. 

The AMG-random agent, on the other hand, achieved surprising results against all top-tier AV products, outperforming both the trained PPO agent and MAB-Malware framework in each scenario. While the random application of our AMG modifications did not produce satisfactory results against pure ML models such as GBDT and MalConv, it did produce evasive AEs against commercially available AVs. The random AMG agent recorded evasion rates ranging from 2.15\% to 34.48\% with an average of 11.65\% among the tested AVs.


\section{Conclusion} \label{sec:conclusion}
In this paper, we presented a black-box evasion attack using reinforcement learning algorithms in the space of PE binaries. To achieve that, we implemented an interactive environment in the OpenAI Gym format for training RL agents. The environment includes a PE file modifier with tested modifications that maximize the preservation of original functionality. Our PE modifier registered the highest validity of modified binaries compared to modifiers from frameworks such as gym-malware or MAB-Malware.

Further, we collected a dataset of 7,000 Windows malware EXEs and experimented with three RL algorithms DQN, PG, and PPO. We optimized the maximum number of modifications and various hyperparameters for each RL agent. Based on the recorded results, the PPO algorithm with $\gamma=0.5$ and $\alpha=0.0001$ achieved the highest evasion rate of 53.84\% against the GBDT classifier while increasing the AE size by 3.57\% on average. Furthermore, we tested the transferability to other malware detectors where our PPO agent achieved an evasion rate of 11.41\% against MalConv and an average evasion rate of 2.31\% against leading AV engines. 

We compared these results to a random agent using the same set of PE modifications as our trained PPO agent and to MAB-Malware, a state-of-the-art generator of malicious AEs. MAB-Malware bypassed the GBDT and MalConv in 76.12\% and 60.1\% of cases, respectively, outperforming both the random and PPO agents. When transferring the generated AEs to real-world AVs, MAB-Malware achieved an average evasion rate of 2.61\%, which is comparable to our trained PPO agent. The random agent, on the other hand, recorded evasion rates ranging from 2.15\% to 34.48\%, with an average of 11.65\%, significantly outperforming both MAB-Malware and our trained PPO agent. These findings show that utilizing GBDT as a substitute model to generate AEs does not result in evasive AEs against real-world AVs and that even leading AVs are vulnerable to well-crafted, yet randomly applied, adversarial perturbations.

While our goal was to implement a functionality-preserving adversarial attack, the next natural step would be to introduce a defense mechanism that could be incorporated into existing detectors. Retraining with generated AEs or a self-contained AE classifier could be used as a defensive technique, but more research must be done in this area.

Additionally, our proposed approach to generating AEs still has room for improvement. To begin, one of our implemented PE modifications (append new import) did not sufficiently preserve the validity of modified files, and more improvements should be made before deploying this operation to other projects. Following that, different modifications or target classifiers could be introduced to further improve the AMG framework.

One of the strong points of our evaluation of the modification validity is its time efficiency, which stems from the usage of hard-coded adversarial perturbation found in RL-based generators of adversarial malware. While predefined modifications are common nowadays, learned perturbations could prevail in the future. As a result, a reliable evaluation of the functionality of PE files should be developed, which could be incorporated into the generation of AEs.

However, our work provides a solid implementation of a reinforcement learning generator working at the level of binary samples while generating functional adversarial malware examples. Additionally, our modifications, agents, and environment setup can be easily extended for future improvements and are freely available to the public.


\backmatter

\bmhead{Acknowledgments}
This work was supported by the Grant Agency of the Czech Technical University in Prague, grant No. SGS23/211/OHK3/3T/18 funded by the MEYS of the Czech Republic and by the OP VVV MEYS funded project CZ.02.1.01/0.0/0.0/16 019/0000765 ``Research Center for Informatics''.

\bibliography{main}

\end{document}